\documentclass[prb,twocolumn,showpacs,aps,floats,floatfix,superscriptaddress]{revtex4}
\usepackage{graphicx,epsfig}

\begin{document}

\title{Scaling relations in the collapse of elastic icosadeltahedral shells under uniform external pressure}
\author{Antonio \v{S}iber}
\email{asiber@ifs.hr}
\affiliation{Institute of Physics, P.O. Box 304, 10001 Zagreb, Croatia}

\author{Rudolf Podgornik}
\affiliation{Department of Theoretical Physics, Jo\v{z}ef Stefan Institute, SI-1000 Ljubljana, Slovenia}
\affiliation{Department of Physics, University of Ljubljana, SI-1000 Ljubljana, Slovenia}

\begin{abstract}
We discuss collapse of icosadeltahedral shells subjected to uniform external pressure. We demonstrate that 
there is a universal collapse pressure curve. The parameter that 
uniquely determines the collapse pressure is shown to be the F\"{o}ppl-von Karman number of the 
non-pressurized shells. Numerical results are interpreted in terms of buckling instabilities of 
spheres and cylinders under hydrostatic pressure. 
\end{abstract}


\maketitle

It has been experimentally demonstrated that microscopic and nanoscopic capsules can collapse 
under external pressure, evaporation of solvent, or point-forcing \cite{Feryrev}. In most of these cases, 
the studied capsules were nearly spherical. Viruses represent an exception as they can take on
different shapes. Even icosahedral viruses that are studied in this article can have very 
different shapes, some of them being nearly perfectly spherical, while others have pronounced 
polyhedral shape and nearly flat faces\cite{Bakerreview}. These viruses can be represented as 
triangulated shapes of spherical topology with an icosa(delta)hedral backbone - this is known as the 
principle of quasiequivlence and forms the basis of the Caspar-Klug classification of 
icosahedral viruses \cite{casparklug}. This also means that viruses can be thought of as 
consisting of clusters of proteins, with $10 (T-1)$ of these clusters consisting of six proteins 
(hexamers) and exactly twelve of them consisting of five proteins (pentamers). Here $T$ is the 
triangulation number of the virus and describes the arrangement and number of protein subunits on the capsid \cite{Lidmar,Siber1}.

The central issue of this work is to explore whether and how the equilibrium shape of an icosadeltahedral shell 
(viral capsid in particular) influences its response to the external mechanical pressure. This should be important for 
empty viral capsids submitted to osmotic pressure of the external solution. The impenetrability of the viral capsid to the osmoticant, such as polyethylene glycol (PEG), gives rise to a mechanical pressure across the capsid shell that compresses it. Experiments along these lines on complete virions (capsids as well as genome) are performed on bacteriophages in PEG bathing solutions \cite{evilevitch1}. Our interest here is not in the pressure dependence of the amount of DNA encapsidated in these viruses \cite{SiberDragar}, but in the response of {\em empty} capsids under mechanical pressure and, most importantly, their collapse. 

\begin{figure}[ht]
\centerline{
\epsfig {file=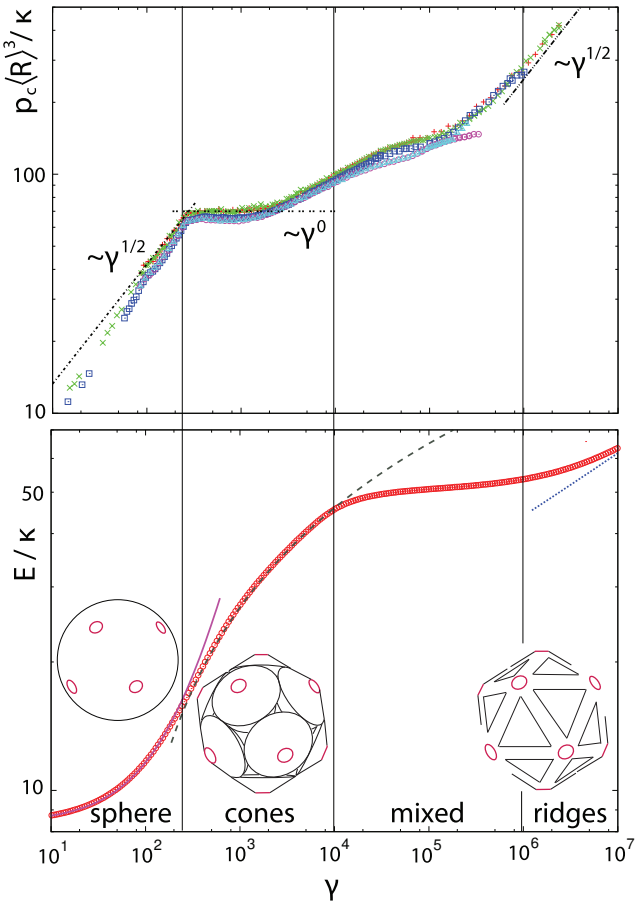,width=8.5cm}
}
\caption{Panel (a): Scaled critical pressures (symbols) of the icosadeltahedral shells with $\epsilon=1$ and $T=441$ ($+$), 
$\epsilon=10$ and $T=441$ ($\times$), $\epsilon=5$ and $T=100$ (squares), $\epsilon=5$ and $T=49$ (circles), and 
$\epsilon=4$ and $T=73$ (triangles) as a function of the FvK number $\gamma$. In these calculations, $u=0.005 a$, $\epsilon$ was kept fixed and $\kappa$ so to produce a variation in $\gamma$. The dashed lines show scalings with $\gamma$ as discussed in the text. Panel (b): The energetics of non-pressurized shells as a function of $\gamma$. The regimes in which the shell can be represented as a sphere, assembly of cones or ridges are indicated. Analytical expressions for the shell energetics are indicated by full (sphere), dashed (cones) and dotted (ridges) lines \cite{Lidmar}.
} 
\label{fig:fig1}
\end{figure}

By applying the  continuum elasticity theory to viral capsids (empty viruses), Lidmar {\em et al} \cite{Lidmar} (LMN in the following) have shown  that their shape can be understood in terms of a {\em single} parameter, the so-called F\"{o}ppl-von K\`{a}rm\`{a}n (FvK) number ($\gamma$) given as
\begin{equation}
\gamma = Y \langle R \rangle ^2 / \kappa.
\end{equation}
Here $Y$, and $\kappa$ are the two-dimensional Young's modulus and the bending rigidity of the viral protein 
sheet, respectively, and $\langle R \rangle$ is the mean radius of the viral capsid. For $\gamma$ smaller than 
about 250, the equilibrium shape of the capsid that minimizes its elastic energy is practically 
a perfect sphere [see Fig. \ref{fig:fig1}b)]. When $250 \lesssim \gamma \lesssim 5000$, a continuous transition in equilibrium shape takes place and the 
capsids assume a more aspherical shape. LMN have termed this the ''buckling 
transition'' since the regions surrounding the pentagonal disclinations (protein pentamers) ''buckle 
out'' from the sphere so that the surface surrounding each of the disclinations is nearly 
conical (it should be understood that the buckling transition in the LMN terminology is not a consequence of 
any external forcing of the capsid but simply results from the minimization of elastic energy of the shell).
In the region $250 \lesssim \gamma \lesssim 10^4$, the capsids can be imagined as a 
union of twelve conical frusta (with apices at the icosahedron vertices) that are fastened together 
at their bases \cite{Tersoff,Siber1}. This approximation provides a good account of the energetics of 
the shells \cite{Lidmar,Siber1,Siber3}. Within the range $10^4 \lesssim \gamma \lesssim 10^6$, the conical description of the shell becomes less satisfactory with regards to the shell energetics [c.f. Fig. 6 of Ref. \onlinecite{Lidmar} and Fig. \ref{fig:fig1}b)] since another creeping transition takes 
place that flattens the icosahedron faces and at the same time sharpening the regions around their edges. 
This effect has been explained by Witten and Li\cite{Li} and Lobkovsky\cite{Lobkovsky1} as originating from the stretching energy along the edges that  becomes prohibitively large as the shell size increases (or as $\gamma$ increases) \cite{Siber3}. The scaling relations that are characteristic for the ridge sharpening are observed \cite{Lidmar} when $\gamma \gtrsim 10^6$, so that 
in the region $10^4 \lesssim \gamma \lesssim 10^6$ neither cones nor ridges can provide adequate representation 
of the shell shape and energetics. Note, however, that the borders of different regimes of the FvK number are 
quite smeared, especially towards the ridge sharpening regime, since the transitions in shell shapes are 
continuous. As the FvK number scales with $\langle R \rangle ^2$, larger viruses should look more polyhedral, and this is indeed a general trend observed in the experimental data (note, however, that the influence of the viral genome on the shell shape has been neglected, so that, strictly speaking, the above consideration should be 
applied only to empty viral capsids \cite{Siber2}).

Since the equilibrium shape of the capsid/shell changes 
with the FvK number, we could {\em a priori} expect that the response of the shells to 
external pressure is also strongly influenced by its FvK number. An intriguing question to 
pose is whether the FvK number uniquely determines the response of the shells to 
external pressure. This has in fact been proven in Ref. \onlinecite{nelsbuckl} where the 
authors considered the susceptibility of shells as a function of their FvK number. That study 
has, however, concentrated on elastic response of the shells to external forcing whereas we shall be 
interested in the collapse of shells. Interestingly, even non-pressurized empty viral shells have 
been found to be prone to collapse in two recent molecular dynamics studies \cite{moldynvir1,moldynvir2}. Our study should be in many respects complementary as it allows a study of the collapse as a function of elastic (coarse grained) parametrization of the protein-protein interactions in viral capsids.

We model the shell as a polyhedron with the icosahedral order whose neighboring vertices 
are connected with springs, so that the stretching energy is given by 
\begin{equation}
H_s = \frac{\epsilon}{2} \sum_{i,j} (|{\bf r}_i - {\bf r}_j| - a)^2,
\end{equation} 
where $\epsilon$ is the spring constant, $a$ is the equilibrium separation of the neighboring 
($i$ and $j$) vertices, and ${\bf r}_i$ is the vector pointing at the $i$-th vertex. The 
bending energy of the shell is given by 
\begin{equation}
H_b = \frac{\tilde{\kappa}}{2} \sum_{I,J} ({\bf n}_I - {\bf n}_J)^2,
\end{equation}
where $I$ and $J$ are two polyhedron faces sharing an edge and ${\bf n}_I$ 
(${\bf n}_J$) is the unit normal of the $I$-th ($J$-th) face. The macroscopic 
elasticity constants of the shell material can be derived from $\epsilon$ and $\kappa$ as 
demonstrated in Ref. \onlinecite{Lidmar}, so that the Young's modulus is 
$Y = 2 \epsilon / \sqrt{3}$, the (mean) bending rigidity $\kappa=\sqrt{3} \tilde{\kappa} / 2$, 
the Gaussian bending rigidity $\kappa_G = - 4 \kappa / 3$, and the Poisson ratio 
$\nu = 1 / 3$. The shape of the shell under external pressure ($p$) is found 
by minimizing the Hamiltonian of the problem, 
\begin{equation}
H = H_s + H_b + p V,
\label{eq:totham}
\end{equation}
where $V=\sum_{I} {\bf r}_{I,1} \cdot ({\bf r}_{I,2} \times {\bf r}_{I,3}) / 6$ is the shell volume, and 
${\bf r}_{I,1}, {\bf r}_{I,2}$, and ${\bf r}_{I,3}$ are the vertices of the $I$-th triangular face in 
the clockwise order. For given shell $T$-number and parameters $a$, $\epsilon$, and $\kappa$, 
the non-pressurized shape of the shell is uniquely determined by its FvK number. Upon increasing pressure, the 
minimum-energy shape of the shell [minimizing $H$ in Eq.(\ref{eq:totham})] changes, i.e. the shell deforms. We perturb the vertices of the thus 
obtained shape by adding a random displacement to each of the vertices, ${\bf r}_i' = {\bf r}_i + u {\bf e}_i$, 
where ${\bf e}_i$ is a random three-dimensional unit vector, different for each vertex, and $u$ is the amplitude 
of the displacement. The perturbed shape can be thought of as a particular conformation excited by (low-, $u \ll a$) temperature fluctuations and the perturbation has in general non-vanishing projections on each of the vibrational 
eigenmodes of the shell (see below). 
We then again minimize the energy of the perturbed shape. For sub-critical pressures, the shape 
equilibrates back to the unperturbed state, but at some critical pressure ($p_c$), the new conformation 
that is adopted by the shell differs from the unperturbed state, i.e a discontinuous transition in the shell shape 
parameters and volume is observed. Note here that we assume that the elastic instability of the shell takes place 
prior to its rupture which is corroborated by the experiments on empty bacteriophage capsids which 
are shown to withstand large forces and indentations by the tip of an atomic force microscope ($\sim$ 6 nm) prior to nonlinear response and possible rupture \cite{ivanovskaPNAS}. 
To relate our results to those previously published for non-pressurized icosadeltahedral 
shells \cite{Lidmar,Siber2} and continuum theories of shell instability \cite{Pogorelov,timoshenkogere,Landau}, we shall be particularly interested in shells with large T-numbers. The results should be therefore directly applicable to viruses with large number of protein subunits, although the numerical simulations can be even more easily performed for shells with small T-numbers.

Deformations of elastic plates subjected to external forces are known to belong to two distinct categories, depending 
on their elastic parameters. It 
has been experimentally demonstrated that deformations of a clamped half-cylindrical surface under 
point forcing \cite{cylinder} and the gravity induced draping of naturally flat, isotropic sheet \cite{draping} can 
be understood in terms of the creation of conical singularities (the so-called d-cones) \cite{conical1}. This means 
that the energy of deformation is dominantly of the {\em bending} type. On the other hand, the crumpling of 
a sheet of paper has been described as dominated by the generation of narrow stretching ridges \cite{Kramer,Wittenrev}. In this case, the energetics of deformation is still dominated by the bending contribution, but the stretching energy becomes of comparable magnitude \cite{Lobkovsky1}. As already discussed, in 
the shapes and energetics of equilibrium non-pressurized shells, both types of deformations can be seen but in different regimes of the FvK number. Thus, we expect that the same two types of deformations should also be seen in the energetics of the shell collapse and this is one of the central issues of this work. LMN have 
demonstrated \cite{Lidmar} that the elastic energy of non-pressurized shells 
scales as $E \propto \kappa f(\gamma)$, where the function $f(\gamma)$ assumes different forms depending upon 
whether the energy of the shell is dominated by stretching or bending (i.e. on the value of $\gamma$). 
A simple analysis suggests that if there is universality in the collapsing pressure, and if it is related to non-pressurized energetics and shape of the shell, it should reveal itself in the scaling $p_c' \rightarrow p_c 
\langle R \rangle ^3 / \kappa$, where $\langle R \rangle$ is the mean radius of the non-pressurized shell. The 
results presented in Fig. \ref{fig:fig1}a) convincingly demonstrate the universality in $p_c'$. 
This figure displays the scaled critical pressures $p_c'$ as a function of FvK number of the shell in the 
non-pressurized state. Note that the thus rescaled collapsing pressures for shells of different elastic 
properties ($Y$ and $\kappa$) and T-numbers (i.e. $\langle R \rangle$ if $a$ is fixed) all fall on the same universal curve which we denote by ${\cal U} (\gamma)$. 
The collapsing pressure is thus given as 
\begin{equation}
p_c = \frac{\kappa}{\langle R \rangle ^3} {\cal U} (\gamma),
\end{equation}
Intriguingly enough, for values of $\gamma$ that should be typical of large viruses ($300 \lesssim \gamma \lesssim 3000$) \cite{Lidmar}, ${\cal U} (\gamma)$ is nearly constant, i.e. ${\cal U}(\gamma) \approx 70$. The measurement of collapsing pressures should thus produce a direct information on $\kappa$.

The collapse of shells as the pressure increases can be understood by following the eigenmodes of shells for a 
given pressure. These are obtained by diagonalization of the matrix of second derivatives of the Hamiltonian in Eq. 
(\ref{eq:totham}), viz. $\partial^2 H / \partial {\bf r}_i \partial {\bf r}_j$. Near the collapsing pressure, 
the frequency of one of the eigenmodes approaches zero indicating the 
instability of the shell with respect to the motion pattern specified by the eigenvector of the critical 
eigenmode. This approach has been successfully applied in the analysis of ridge buckling instability 
\cite{diDonna}. We have found that the eigenvector patterns of critical eigenmodes change with $\gamma$, so that 
different shell motions lead to its instability, depending on $\gamma$. One could proceed further with 
this analysis but a lot of insight in the 
numerically obtained results can be obtained by applying existing analytical results for the critical 
hydrostatic pressures leading to buckling of spheres \cite{Pogorelov,Landau,timoshenkogere} and cylinders \cite{timoshenkogere}. 
Namely, although the shapes of both pressurized and unpressurized shells are fairly complex, their surface can be 
separated in regions that locally resemble spheres (two finite radii of curvature) or cylinders 
(one radius of curvature tending to infinity), depending on the FvK number. We thus expect that the scaling of 
critical pressures for spheres and cylinders should be seen in our numerical data in different regions 
of $\gamma$. For FvK number smaller 
than about 250, 
the equilibrium shape of the unpressurized 
shell is practically a perfect sphere \cite{Lidmar,Siber2}. Therefore, the critical pressures should be proportional to 
\cite{Pogorelov,timoshenkogere,Landau}
\begin{equation}
p_c^{{\rm sphere}} \propto \frac{\kappa \sqrt{\gamma}}{\langle R \rangle ^3},
\label{eq:crit_sphere}
\end{equation}
which is an expression for collapse pressure of thin spheres. 
This is exactly what we find from our numerical results in the region $90 \lesssim \gamma \lesssim 250$ 
as can be seen by comparison of a function proportional 
to $\gamma ^{1/2}$ with the data. A deviation from this behavior occurs for $\gamma < 90$, but note that 
for a shell made of uniform elastic material $\gamma = 12 (1-\nu^2)(R/d)^2$, where $d$ is an effective thickness of 
the shell. Thus, for $\gamma \lesssim 90$, $R/d$ is no longer negligible and an expression for thin shells 
in Eq. (\ref{eq:crit_sphere}) becomes inapplicable.
For shells that underwent the buckling transition in their non-pressurized state, the shell surface is nearly 
conical around the pentagonal disclinations (protein pentamers) i.e. locally cylindrical, 
so that the critical pressure should be proportional to \cite{timoshenkogere}
\begin{equation}
p_c^{{\rm cylinder}} \propto \frac{\kappa}{R_{cyl}^3},
\label{eq:cylind}
\end{equation}
where $R_{cyl}$ is the maximum radius of the cylindrical regions in the shell (this suggest that the critical points 
of collapse are located in the middle of icosahedron edges). The maximum radii of the cones 
are $R_{cyl} \sim \langle R \rangle$, so that the 
critical pressures should be independent of $\gamma$, which is also obtained in our numerical results 
for $300 \lesssim \gamma \lesssim 3000$ (i.e. ${\cal U}(\gamma) \propto {\rm const.}$). 
As the FvK number increases, the shell shape just prior to collapsing shows flattening of the faces 
and concentration of curvatures along the ridges, so that the radius of cylindrical regions 
immediately before collapse notably decreases from its non-pressurized value as a result of applied pressure. 
This can also be 
seen as an increase in the collapse pressure, i.e. its deviation from the constant value predicted by Eq. (\ref{eq:cylind}). 
Note also that the description of shells in terms of 
cones becomes inaccurate when $10^4 \lesssim \gamma \lesssim 10^6$ [see Fig. \ref{fig:fig1}b)] and neither ridges nor 
cones provide an adequate description of the shell shape in this region of FvK number since the curvature 
along ridges scales in a complicated way with $R$ and $\gamma$. When $\gamma \gtrsim 10^6$, the 
ridge-sharpening transition gradually takes place, and the characteristic curvature radii of (non-pressurized) 
ridges scale as \cite{Lobkovsky1}
\begin{equation}
R_{cyl} \propto \langle R \rangle \gamma ^{-\frac{1}{6}},
\end{equation}
which means that the critical pressures for shell collapse should scale as 
\begin{equation}
p_c \propto \frac{\kappa}{R_{cyl}^3} \propto \frac{\kappa \sqrt{\gamma}}{\langle R \rangle ^3},
\label{eq:crit_ridges}
\end{equation}
i.e. functionally the same as for shells that are spherical in their non-pressurized state 
[$\gamma \lesssim 250$, see Eq. (\ref{eq:crit_sphere})]. 
The scaling of $p_c$ with $\gamma ^{1/2}$ in this regime (${\cal U} (\gamma) \propto \gamma ^{1/2}$) 
is again strongly suggested by our numerical results. A detailed exploration of this region of large FvK number 
requires numerical studies of very large shells (large $T$-numbers) since the lower bound on radii of 
ridge curvatures is set by $a$. This slows down the calculations, but from our results we estimate that 
$T=441$ shells should be a reliable model for continuum shells to about $\gamma \sim 10^8$. The region of 
such large $\gamma$ is not important for viruses but may be of importance for other shell structures with 
icosahedral order such as e.g. self-assembled hollow icosahedra in salt-free catanionic solutions 
\cite{gianticosahedra}.

We have confirmed that the main features of our numerical results on the collapse of elastic icosadeltahedral shells under uniform external pressures can be understood on the basis of buckling instability pressures that 
are characteristic of spherical and cylindrical surfaces. Note that the scaling obtained in our results 
is also predicted by Eqs. (\ref{eq:crit_sphere}), (\ref{eq:cylind}), and (\ref{eq:crit_ridges}).
The knowledge of bending rigidity of viral capsids is required in order to apply our results to viruses. This 
quantity is unfortunately not known precisely, but can be estimated to be of the order of $\kappa \sim 40 k_B T$
from a previous study of bacteriophage shells under internal pressure generated by DNA \cite{Siber2} (a 
lower value of about 10 - 15 $k_B T$ has been found in Ref. \onlinecite{Bruinelast}). Assuming a 
radius of about $\langle R \rangle \sim$ 30 nm that is typical for e.g. $\lambda$-bacteriophage produces 
collapsing pressures in the regime of FvK numbers that are typical for large viruses of about $p_c \sim 5$ atm. The 
pressures of this order of magnitude can be easily achieved in experiments \cite{evilevitch1}, but we again 
stress that a test of our prediction would require exposing {\em empty} viruses to osmotic pressure. In 
our elastic model we have assumed that the spontaneous curvature of the viral shell is zero. While it is 
still a matter of debate whether the protein coatings of viruses do indeed have some non-vanishing spontaneous curvature, this is suggested by a recent study \cite{SiberPodgornik}. This may also have some influence on the precise value of collapse pressures.

A\v{S} acknowledges support by the Ministry of Science, Education and Sports of Republic of Croatia through the 
Research Project No. 035-0352828-2837. RP acknowledges support by the Agency for Research and Development 
of Slovenia (Grants P1-0055, Z1-7171 and L2-7080).

\end{document}